# The Chemical Evolution of Damped Lyman $\alpha$ Galaxies

Max Pettini[1], David L. King[1], Linda J. Smith[2] and Richard W. Hunstead[3]

[1] Royal Greenwich Observatory, Madingley Road, Cambridge, CB3 0EZ, UK
[2] Department of Physics and Astronomy, University College London, Gower Street, London WC1E 6BT, UK
[3] School of Physics, University of Sydney, NSW 2006, Australia

**Abstract.** Measurements of element abundances in damped Lyman $\alpha$ systems are providing new means to investigate the chemical evolution of galaxies, particularly at early times. We review progress in this area, concentrating on recent efforts to extend the range of existing surveys to both higher and lower redshifts.

## 1 Introduction

One has only to glance at the programme of this workshop, and compare it with that of the last QSO Absorption Line meeting which took place at the Space Telescope Science Institute in April 1987, to realise the increasing amount of attention which damped Lyman $\alpha$ systems have attracted in the last few years. I would like to echo the words by Ray Weymann earlier today, in recognising that this is largely due to the efforts by Artie Wolfe and his collaborators who, in a series of large-scale studies, identified several lines of evidence to suggest that in this class of absorbers we may well be seeing the high-redshift progenitors of galaxies like our own. If we accept this as a working assumption, we can then bring the full power of QSO absorption line spectroscopy to bear in determining many of the physical properties of galaxies caught, as it were, in their infancy.

## 2 Heavy Element Abundances at Redshift $z = 2$

My collaborators and I have been involved for some years now in a programme aimed at studying the chemical evolution of high-redshift galaxies through measurements in particular of the metallicity, dust content, and star formation in the gas giving rise to damped Lyman $\alpha$ systems. The first stage of our survey was completed and published earlier this year (Pettini et al. 1994);

I shall use those results as a starting point for my talk today, which concentrates on recent work to extend the survey over a wider redshift interval. The filled circles in Fig.1 show the abundances of zinc measured in 17 damped Lyman $\alpha$ systems by Pettini et al. (1994). Pettini, Boksenberg and Hunstead (1989) first drew attention to several advantages in using Zn as a tracer of metallicity; as these are now widely recognised, we only need to summarise them briefly here. Zn is among the few heavy elements which show little affinity for dust; this, coupled with the fact that it is mostly singly ionised in H I regions, makes it likely that the 'missing' fraction of Zn – either in solid form or in unobserved ionisation stages – is small. For most known damped Lyman $\alpha$ systems, multiplet 1 of Zn II, $\lambda\lambda 2025, 2062$ is redshifted into a region of the optical spectrum which can be readily observed. Furthermore, Zn is a relatively rare element ($[Zn/H]_{\odot} = 3.8 \times 10^{-8}$; Aller 1987); consequently the doublet lines are usually sufficiently weak that line saturation can be assessed – and the column density deduced – with an accuracy which is adequate for the present purposes. In terms of galactic chemical evolution, the metallicity measured by the Zn abundance is analogous to that traced by Fe; the nucleosynthetic origins of these two elements presumably have much in common, since [Zn/Fe] shows no departure from solar values in Galactic stars of all metallicities, down to the lowest values measured.

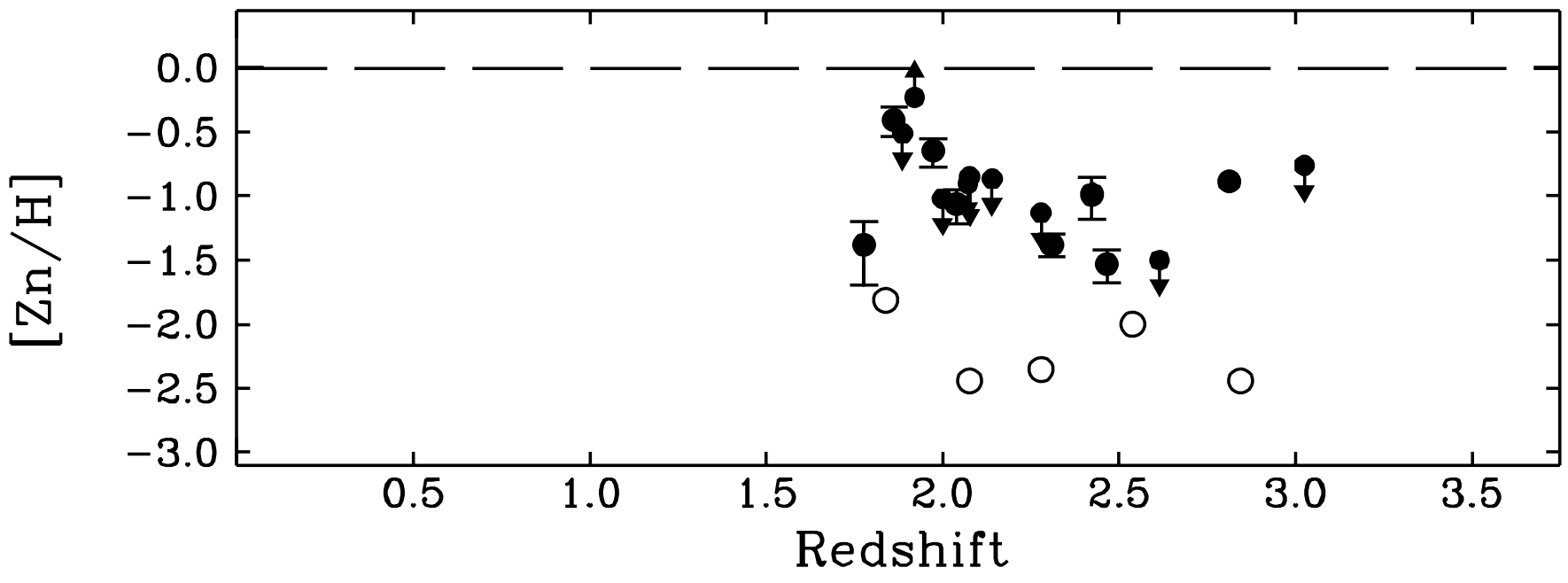

**Fig. 1.** Values of the abundance of Zn in 17 damped Lyman $\alpha$ systems reported by Pettini et al. 1994 (filled circles). [Zn/H] is plotted on a logarithmic scale relative to the solar value; thus the broken line at 0.0 corresponds to the solar abundance and $[Zn/H] = -1.0$ indicates an underabundance of Zn by a factor of 10. Downward pointing arrows are upper limits appropriate to cases where the Zn II lines have not been detected. Also shown for comparison are metal abundances measured from very high resolution echelle observations of damped Lyman $\alpha$ systems (open circles), as follows (in order of increasing redshift): [O/H] in the $z_{\rm abs} = 1.83856$ system in Q1101−264 (Pettini et al. 1992); [Si/H] in the $z_{\rm abs} = 2.07623$ system in Q2206−199 (Pettini and Hunstead 1990); [Fe/H] in the $z_{\rm abs} = 2.27936$ system in Q2348−147 (Pettini, Lipman and Hunstead 1995); [Fe/H] in the $z_{\rm abs} = 2.53788$ system in Q2344+124 (Lipman, Pettini and Hunstead 1995); and [Fe/H] in the $z_{\rm abs} = 2.8443$ system in Q1946+769 (Lu et al. 1995) . Two of the five open circles (the points for Q2206−199 and Q2348−147) have corresponding [Zn/H] upper limits.

Returning to Fig.1, two main results are indicated. First, at redshifts between $z \simeq 2$ and 3, the galaxies giving rise to the absorption are mostly chemically young systems – most of the points in the plot lie well below the solar [Zn/H]. As pointed out by Lanzetta (1993), if the damped Lyman $\alpha$ lines account for most of the baryons at these redshifts, then the column density-weighted metallicity we find, $Z_{\rm DLA} = 1/10 Z_{\odot}$, can be interpreted as the typical 'cosmic' metallicity reached by the universe $\sim$ 13 Gyr ago ($H_0 = 50$ km s$^{-1}$ Mpc$^{-1}$, $q_0 = 0.01$).

Second, there is a large dispersion in the degree of metal enrichment attained by different galaxies at essentially the same epoch. About half of the filled circles in Fig.1 are upper limits, corresponding to non-detections of the Zn II doublet; in some of these cases we *know* that the true abundances lie well below the limits of the Zn II observations. For metallicities $Z_{\rm DLA} < 1/50 Z_{\odot}$, the Zn II lines become vanishingly small but, on the other hand, we enter a regime where the ultraviolet resonance lines of the more abundant astrophysical elements become accessible to abundance studies through high-resolution observations with echelle spectrographs. Five such measurements are included as open circles in Fig.1 . There may be uncertainties of 0.3-0.5 dex in comparing these data with the values of [Zn/H], since some of the elements in question (Si and O in particular) may be present in non-solar proportions relative to Zn. Nevertheless, the inclusion of these points in the figure does highlight the fact that at $z \simeq 2$ damped Lyman $\alpha$ systems span more than two orders of magnitude in metallicity and that systems with $Z_{\rm DLA} < 1/100 Z_{\odot}$ are not uncommon.

Such a wide range is unlikely to be due to radial abundance gradients similar to those seen in present-day spiral galaxies. Simulations show that, within the inner regions of galaxies where the column density of neutral gas is sufficiently high to produce a damped Lyman $\alpha$ line ($N({\rm H}^0) \geq 2 \times 10^{20}$ cm$^{-2}$) randomly placed sight-lines would on average sample an abundance spread of only a factor $\approx 2$ . While it is possible that the interstellar medium at these high redshifts is poorly mixed, the observations suggest that the process of chemical enrichment started at different times and probably proceeded at different rates in the galaxies picked out by the damped systems.

## 3 Extension to Higher Redshifts

The results in Fig.1 have been interpreted by some (e.g. Wolfe 1993) as evidence for a rapid build-up of metals in the universe between $z = 3$ and 2. Such an effect may go hand-in-hand – at least qualitatively – with the significant consumption of gas apparently indicated by the evolution, over approximately the same redshift interval, of the H I distribution of damped Lyman $\alpha$ lines (Lanzetta, Wolfe and Turnshek 1995). In our view, however, published data are insufficient to measure with confidence trends in either metallicity *or* H I content with redshift. Indeed, the recent finding that a large sample of very high redshift ($z_{\rm em} > 4$) QSOs includes significantly fewer systems with $N(\rm H^0) > 10^{21}$ cm$^{-2}$ than expected (Storrie-Lombardi et al. 1995) suggests that the H I redshift evolution may have been overestimated. As far as the Zn measurements in Fig.1 are concerned, while it is true that near-solar abundances are found only at $z \simeq 2$, the redshift distribution of the data in Fig.1 is too uneven to assess the significance of this result.

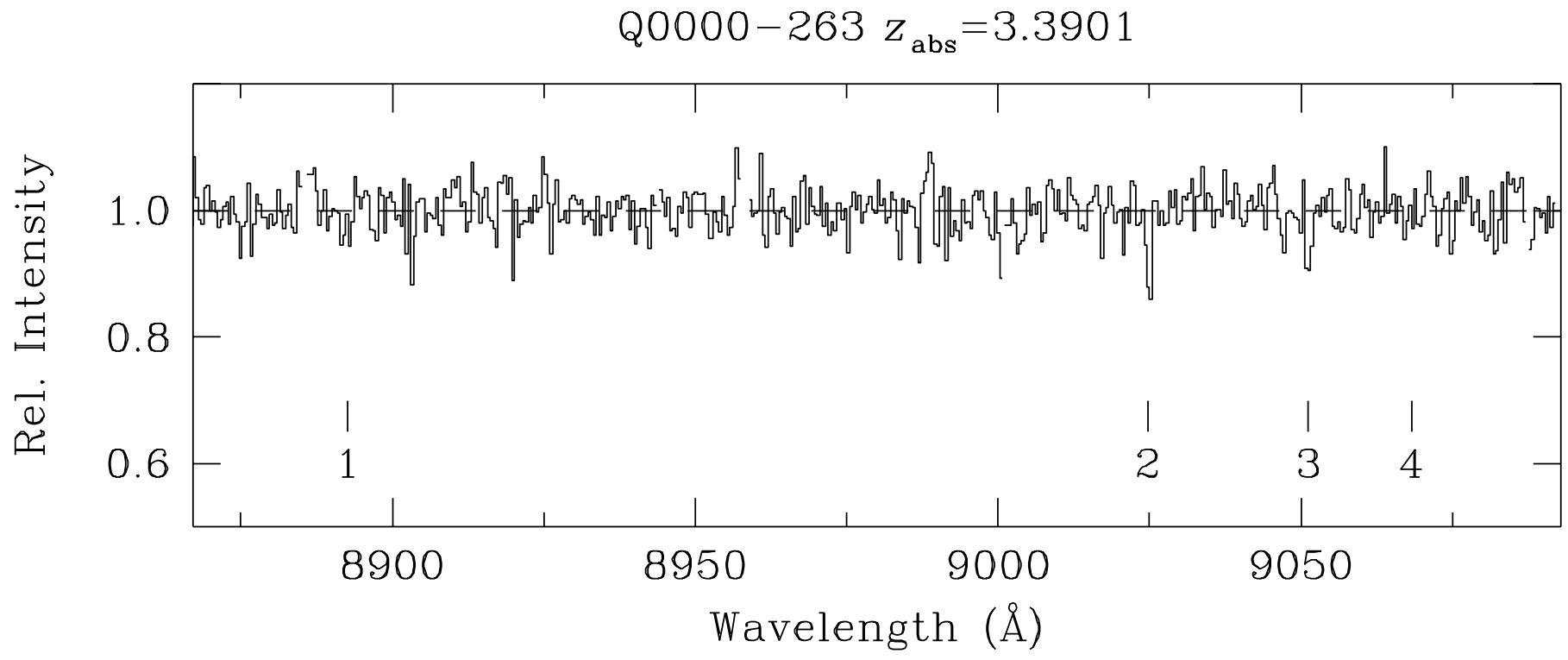

**Fig. 2.** Portion of the AAT spectrum of Q0000−263 encompassing the redshifted Zn II and Cr II lines in the $z_{\rm abs} = 3.3901$ absorption system. The tick marks indicate the expected positions of the absorption features, whether they have been detected or not. Line 1: Zn II $\lambda$2025.484; line 2: Cr II $\lambda$2055.596; line 3: Cr II $\lambda$2061.575+ Zn II $\lambda$2062.003 (blended); and line 4: Cr II $\lambda$2065.501. The spectrum shown here is the sum of many exposures with a Tektronix CCD, amounting to a total integration time of 58 000 s; the resolution is 1.1 Å *FWHM* and the final S/N $\simeq$ 30 .

To remedy the situation, we have extended the survey in 1994 with observations of an additional 7 damped Lyman $\alpha$ systems, mostly at redshifts $z_{\rm abs} \geq 2.5$ . The full sample for which Zn abundance determinations (or upper limits) are now available consists of 24 damped systems in 20 QSOs, and includes more than one third of the total number of confirmed damped absorbers. The new data were obtained during a series of observing runs at the 3.9 m Anglo-Australian telescope at Siding Spring Observatory, Australia, and the 4.2 m William Herschel telescope at La Palma, Canary Islands. The instrumental set-ups were similar to those used by Pettini et al. (1994); the spectral resolution ranged from 0.7 to 1.2 Å *FWHM*.

The observations of Q0000−263, a bright, high-redshift ($z_{\rm em} = 4.111$) QSO, reproduced in Fig.2, are among the most sensitive in the survey. By adding together several nights' data, we achieved a moderately high signal-to-noise ratio (S/N $\simeq$ 30) in the difficult region near 9000 Å, to where the Zn II and Cr II lines associated with the $z_{\rm abs} = 3.3901$ damped system are redshifted. No Zn II absorption is detected, implying a $3\sigma$ upper limit [Zn/H] $< -1.76$, or less than 1/60 of solar! This limit is $\sim 5$ times more sensitive than that placed by Savaglio, D'Odorico and Moller (1994). Interestingly, we have a clear detection of Cr II absorption; we deduce [Cr/H] = −2.46, nearly 300 times below the solar abundance. This is one of the few examples now known where the Cr II lines are stronger than those of Zn II, implying little – if any – depletion of Cr onto dust grains. In Pettini et al. (1994) we predicted that such cases would be discovered at the lowest metallicities, on the basis of an apparent trend in our data of increasing [Cr/Zn] with decreasing $Z_{\rm DLA}$ .

In general, however, the observations we attempted are at the limit of what can be accomplished with 4 m-class telescopes, due to the decreasing response of CCDs and increasing sky emission at wavelengths longwards of $\sim 7000$ Å. Consequently, we were not always able to reach as high a sensitivity to the Zn II lines as we would have wished.

As can be seen from Fig.3, we have found no new detections of Zn II at $z_{\rm abs} \geq 2.5$. The only system showing measurable amounts of Zn at these redshifts remains the $z_{\rm abs} = 2.8122$ absorber in Q0528−250 (Meyer, Welty and York 1989) which is at a higher redshift than the QSO itself and is probably atypical of the sample as a whole in other ways (Moller and Warren 1993). Despite their limitations, the available data therefore seem to suggest that abundances at $z \geq 2.5$ *are* generally lower than 1/10 of solar, the typical value at $z \simeq 2$. *All* damped Lyman $\alpha$ systems sampled at $z \geq 2.5$ are metal-poor, while at $z \simeq 2$ at least some galaxies have apparently attained a significant degree of metal-enrichment, comparable to that of spiral galaxies today. It may well be, then, that the interval between $z = 3$ and 2 does signal the epoch of galaxy formation, if we take the term to indicate the period when the first major episodes of star formation took place in galaxies.

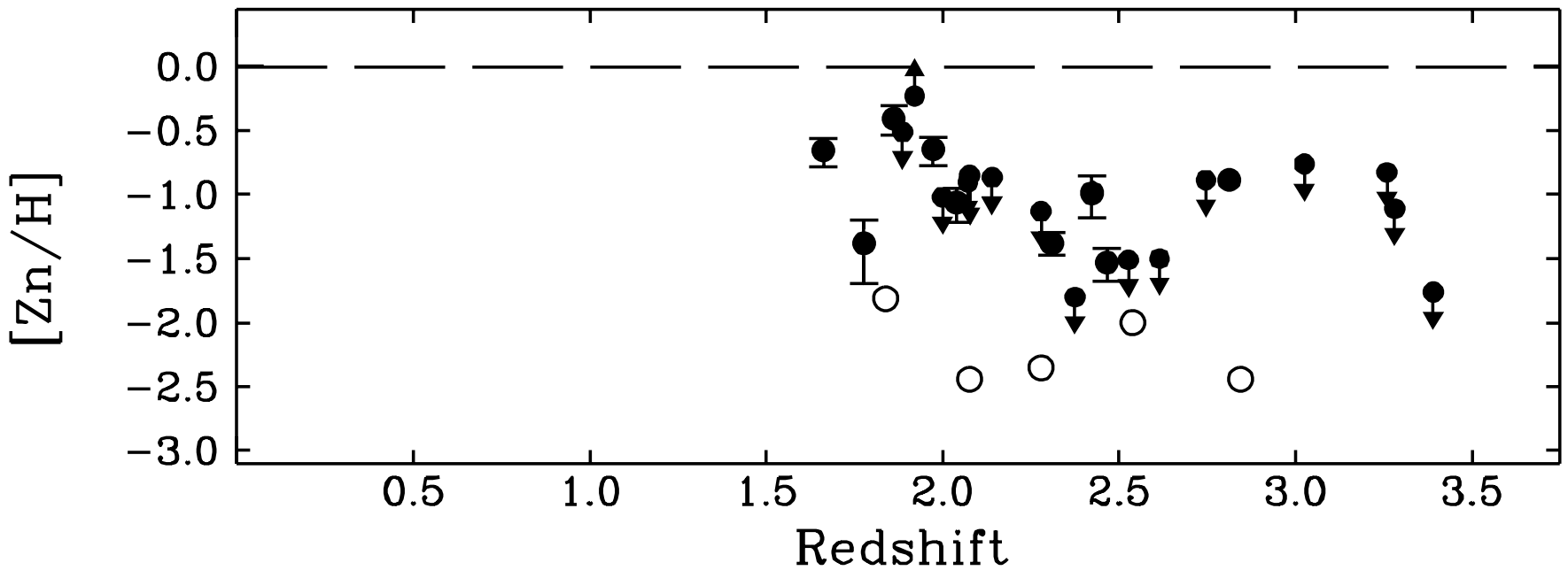

**Fig. 3.** Same as Fig.1, but with the addition of seven new measurements of [Zn/H] obtained in 1994, mostly for systems at $z_{\rm abs} \geq 2.5$.

In this scenario the damped system in Q0000−263 may be a good example of a primeval galaxy. The absorber has been imaged in the stellar ultraviolet continuum near the Lyman limit by Steidel and Hamilton (1992) and it appears to be a luminous galaxy ($L_{\rm B} \approx 3L_*$) of dimensions $10-20\ h^{-1}$ kpc.

Knots which may be sites of star formation are seen in a recent *Hubble Space Telescope* image (Giavalisco 1995, private communication). And yet the interstellar medium of this galaxy – at least in the region which happens to lie in our line of sight to Q0000−263 – has apparently undergone little, if any, chemical enrichment. The metallicity deduced from Cr II and confirmed by observations of other elements by Vladilo et al. (1995), $Z_{\rm DLA} = -2.5$, is the same as that which apparently applies to clouds in the Lyman $\alpha$ forest, as recently discovered by Tytler (1995). One possible interpretation is that this value may represent a 'base level' of metallicity on which the process of galactic chemical evolution subsequently builds up, although some stars in our Galaxy must have formed from more pristine gas, since stars with abundances as low as $Z \simeq -4$ are known (e.g. McWilliam et al. 1995).

Before moving on, we point out that, if the Lyman $\alpha$ forest does indeed trace an intergalactic medium which at $z \simeq 3$ has been already 'polluted' with heavy elements, abundance measurements of a wide complement of elements in damped systems with metallicities as low as $Z_{\rm DLA} = -2.5$ may offer vital chemical clues to the stellar populations which produced this initial enrichment of the universe at early epochs.

## 4 Metal Abundances at Intermediate Redshifts

The next step in constructing a full picture of the chemical evolution of galaxies is to follow the build-up of metals to recent epochs. The difficulty here is not so much in detecting the Zn II lines (with modern blue-sensitive CCDs it is relatively straightforward to search for these down to redshifts $z_{\rm abs} \simeq 0.6$), but rather in identifying even a modest sample of damped Lyman $\alpha$ systems. Cosmological effects, combined with the intrinsic evolution of the H I content, result in significantly smaller numbers of candidates than at higher redshifts (Lanzetta, Turnshek and Sandoval 1993; Bahcall et al. 1993); these difficulties are compounded by the fact that *HST* observations are then required to confirm the candidates and measure $N({\rm H}^0)$ .

Nevertheless there is a strong incentive to extend this work to $z < 1$, particularly given the spectacular success in identifying QSO absorbers at these redshifts by direct imaging (Steidel 1995). The morphological information provided by the images, together with the physical properties of the absorbing gas deduced from QSO absorption line spectroscopy, form a particularly powerful combination for unravelling the nature and evolutionary status of the absorbing galaxies.

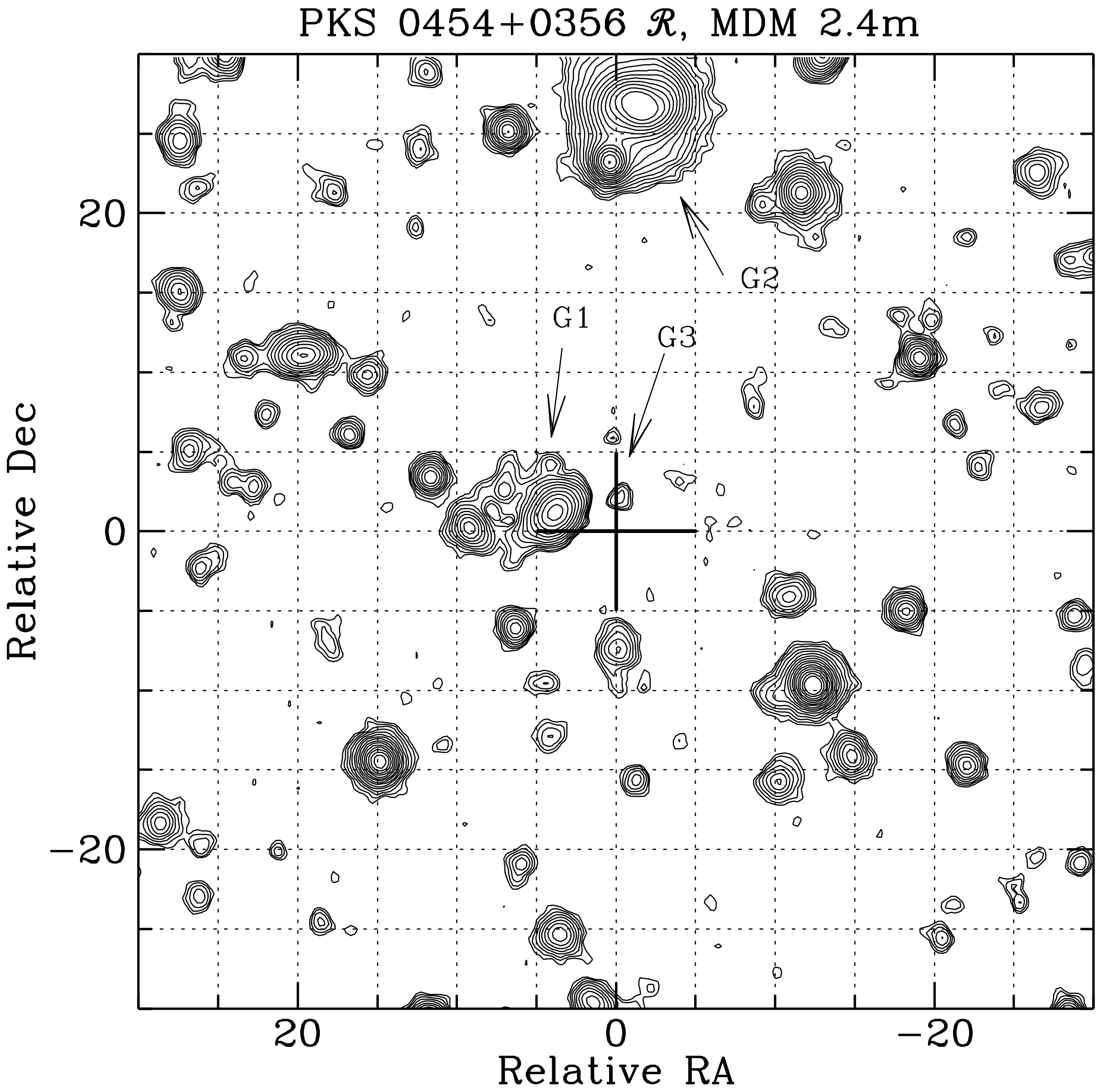

**Fig. 4.** (*Reproduced from Steidel et al. 1995*). $R$ band image of the field of PKS 0454+039 obtained with a 7600 s-long exposure at the 2.4 m Hiltner telescope at the MDM Observatory on Kitt Peak. The *FWHM* of the seeing profile on the final co-added image is 0.89 arcsec. The image of the QSO has been subtracted by modelling the PSF of bright stars in the field. The galaxies labelled G1 and G2 are at redshifts $z = 0.072$ and 0.201 respectively. G3 is the most likely candidate for the damped absorber at $z_{\rm abs} = 0.8596$ .

To date there are only three measurements of metallicity in damped systems at intermediate redshifts, at $z_{\rm abs} = 0.6922$ in 3C 286 (Meyer and York 1992), $z_{\rm abs} = 0.8596$ in PKS 0454+039 (Steidel et al. 1995), and $z_{\rm abs} = 1.3718$ in Q0935+417 (Meyer, Lanzetta and Wolfe 1995). Surprisingly, in all three cases the abundances deduced are approximately one order of magnitude below solar and *not* significantly higher than the typical metallicity at $z = 2$ ([Zn/H] $= -1.2$, $-0.9$ and $-0.7$ in 3C 286, PKS 0454+039 and Q0935+417 respectively. The last two values of [Zn/H] have been increased by 0.14 over those reported by the original authors for consistency with the values of oscillator strength and solar abundance adopted in the rest of this survey).

The galaxies likely to be responsible for the absorption in 3C 286 and PKS 0454+039 have been identified from deep CCD images obtained in good seeing conditions and their characteristics found to be consistent with the low metal abundances measured. The absorber in front of 3C 286 is probably a luminous galaxy ($M_{\rm B} = -20.8$, or $\approx 0.8L*$), but of *low surface brightness* (Steidel et al. 1994). These authors deduced a peak surface brightness in the rest-frame B band $\mu_{\rm B}(0) \simeq 23.6$ mag arcsec$^{-2}$, close to the median value of the sample of LSB galaxies studied by McGaugh and Bothun (1994) and significantly lower than the Freeman value for 'normal' spirals, $\mu_{\rm B}(0) \simeq 21.65 \pm 0.3$ mag arcsec$^{-2}$. Chemical enrichment probably proceeds at a slower pace in LSB galaxies which, even at the present epoch, have low metal abundances (McGaugh 1994); it is not surprising, then, to find [Zn/H] $= -1.2$ in one such galaxy at $z = 0.6922$.

The field of PKS 0454+039 is shown in Fig.4. The most likely candidate for the $z_{\rm abs} = 0.8596$ absorber is the object labelled G3. If the identification is correct, the galaxy is located $\approx 9h^{-1}$ kpc ($q_0 = 0.5$) from the QSO sight-line and has an absolute magnitude $M_{\rm B} = -19.0$ ($\approx 0.15L*$). Again, the fact that this is a relatively underluminous galaxy provides a plausible explanation for the low metallicity deduced by Steidel et al. (1995).

## 5 Bringing it All Together...

We bring the above results together in Fig.5, where we plot the full set of abundance measurements on a timescale compatible with stellar ages, and compare them with Fe abundances of F and G dwarf stars in the disk of the Milky Way from the landmark paper by Edvardsson et al. (1993). The differences are very obvious. At $z > 2$, corresponding to lookback times of more than $\sim 12.5$ Gyr, both the *distribution* and the *mean* of the abundances measured in damped systems resemble more closely the values found in stars in the halo, rather than the disk, of our Galaxy (Pettini et al. 1994). Presumably at this epoch most galaxies had not yet collapsed to form a thin disk. Furthermore, the only two galaxies observed at a lookback time approaching the age of the Sun ($\sim 5$ Gyr) are evidently on very different evolutionary tracks from that of the Milky Way.

Nevertheless, these findings are not necessarily inconsistent with our working assumption that in the damped Lyman $\alpha$ systems we are seeing the progenitors of present-day spiral galaxies. The extensive survey of Mg II absorbers (of which the damped systems are presumably a subset) by Steidel and collaborators (Steidel, Dickinson and Persson 1994; Steidel 1995) has led to the first determination of the luminosity function of field galaxies picked out by absorption cross-section at $z \simeq 0.6$. The luminosity function is Gaussian, centred near $M_B \simeq -20.5$ (for $H_0 = 50$ km s$^{-1}$ Mpc$^{-1}$). The key question, which can only be addressed with a larger sample of damped systems at intermediate redshift, is whether the distribution of metallicities can be understood in terms of the absorbers' luminosity function, appropriately modified by the scaling of absorption cross-section with luminosity (Steidel, Dickinson and Persson 1994), and the present-day dependence of metallicity on galaxy luminosity (Skillman, Kennicutt and Hodge 1989).

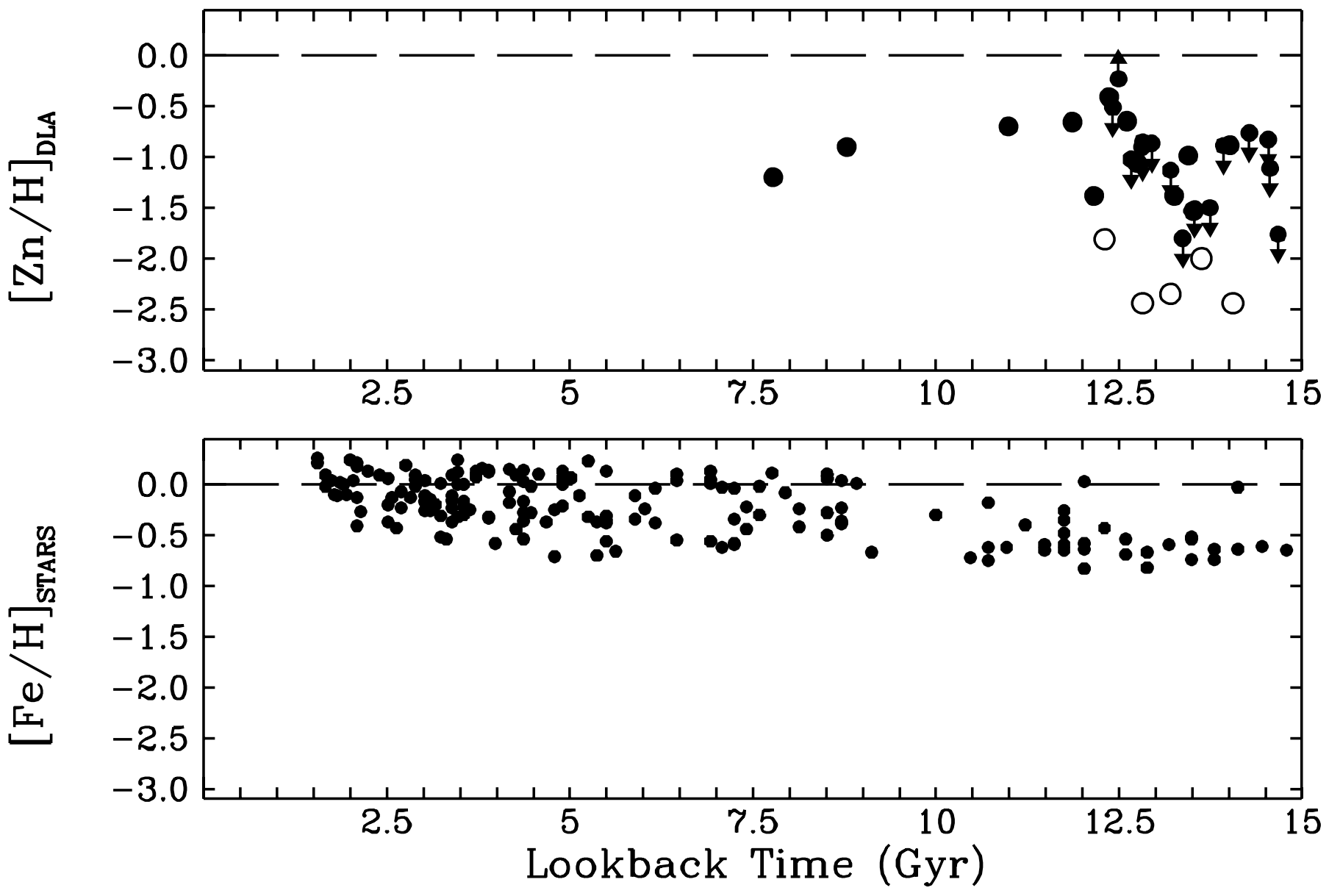

**Fig. 5.** *Top Panel:* Available measurements of metallicity in damped Lyman $\alpha$ galaxies are plotted as a function of lookback time for $H_0 = 50$ km s$^{-1}$ Mpc and $q_0 = 0.01$, this being the set of cosmological parameters which is least discordant with stellar ages. The symbols have the same meaning as in Fig.1 . *Bottom Panel:* Metallicities of 182 disk stars with measured iron abundances and ages from the large-scale study by Edvardsson et al. (1993).

Another factor contributing to the differences evident in Fig.5 between the chemical evolution of the damped systems and that of the Milky Way disk may be the increasing bias introduced by dust as the metal content of the universe grows with time (Fall 1995). In these models, our view of distant galaxies becomes progressively more skewed in favour of metal-poor systems with decreasing redshift. QSOs which lie behind metal-rich – and presumably dusty – galaxies are reddened by amounts which, while too small to hide the QSOs completely, are nevertheless sufficient to exclude them from magnitude limited samples such as those from which the current damped Lyman $\alpha$ surveys are drawn.

To conclude, it is clear that the technique of QSO absorption line spectroscopy has matured significantly in recent years, to the point where it can be used effectively to probe the chemical evolution of galaxies over cosmologically long timescales. It is a powerful new tool at our disposal in an area of study which until now has been based mainly on observations of different stellar populations in the Milky Way and of H II regions in nearby galaxies. Several aspects of the early stages of chemical enrichment in galaxies are already emerging from this work and, with the new generation of 8-10 m class telescopes, we can confidently expect significant progress towards building a full picture.